\documentclass[aps,prb, 10pt,twocolumn,floatfix, footinbib,superscriptaddress]{revtex4-1}

\usepackage{graphicx, color}
\usepackage{amsmath, amsfonts, amssymb, mathrsfs, dsfont}
\usepackage[usenames,dvipsnames, pdftex]{xcolor}
\usepackage{ulem} 

\usepackage[mathscr]{eucal}

\usepackage[colorlinks, breaklinks, 
            linkcolor=OrangeRed,
            citecolor=RoyalBlue,
            urlcolor=RoyalBlue]{hyperref}
            
\usepackage[all]{hypcap} 

\newcommand{\Le}{\left}
\newcommand{\Ri}{\right}

\newcommand{\f}{\frac}

\newcommand{\dg}{\dagger}
\newcommand{\ra}{\rangle}
\newcommand{\la}{\langle}

\newcommand{\eq}[1]{\begin{align}#1\end{align}}

\newcommand{\msr}{\mathscr}

\begin{document}

\title{Anomalous diffusion in particle-hole symmetric many-body localized systems}
\author{Giuseppe De Tomasi}
\affiliation{T.C.M. Group, Cavendish Laboratory, JJ Thomson Avenue, Cambridge CB3 0HE, United Kingdom}
\affiliation{Department of Physics, T42, Technische Universit\"at M\"unchen, James-Franck-Stra{\ss}e 1, D-85748, Garching, Germany}
\author{Daniele Trapin}
\affiliation{Max-Planck-Institut f\"ur Physik Komplexer Systeme, N\"othnitzer Stra{\ss}e 38, 01187-Dresden, Germany}
\author{Markus Heyl}
\affiliation{Max-Planck-Institut f\"ur Physik Komplexer Systeme, N\"othnitzer Stra{\ss}e 38, 01187-Dresden, Germany}
\author{Soumya Bera}
\affiliation{Department of Physics, Indian Institute of Technology Bombay, Mumbai 400076, India}

\begin{abstract}
In this work we probe the dynamics of the particle-hole symmetric many-body localized (MBL) phase.
%
We provide numerical evidence that it can be characterized by an algebraic propagation of both entanglement and charge, unlike in the conventional MBL case.
We explain the mechanism of this anomalous diffusion through a formation of bound states, which coherently propagate via long-range resonances. 
By projecting onto the two-particle sector of the particle-hole symmetric model, we show that the formation and observed subdiffusive dynamics is a consequence of an interplay between symmetry and interactions. 
\end{abstract}

\maketitle

\section{Introduction}
%
Symmetry and dimensionality both play a substantial role in the physics of single-particle Anderson localization~\cite{Mirlin00}.
For instance, in one- and two-dimensional systems without the presence of any specific symmetry all single-particle wave-functions are exponentially localized by any infinitesimal amount of disorder~\cite{mott1961theory, Goldshtein1977, Lucci79, Mirlin00}.
On the other hand, the presence of particular symmetries can alter the situation and can give rise to electronic states that are not exponentially localized. 
%
%
%
%
%
%
In low-dimensional systems, critically delocalized states appear due to protection by chiral symmetry. 
%
%
%
Additionally, in one dimension the presence of chiral symmetry leads to a singularity in the density of states at zero energy, which is known as Dyson singularity~\cite{Dyson53,Weissmann_1975, Theo76, Eggarter78, Fleishman_1977,Souk81,Brower2000, Bush_1975, GDT_2016, Mirlin_2019}. 
Such a model in one dimension is realized by having off-diagonal or bond disorder instead of random chemical potentials.
Here, the single-particle eigenstates with energy close to the singularity are quasi-localized and the ground state belongs to a random singlets phase~\cite{Fisher94, Theo76, Eggarter78, Ziman82, Bush_1975, GDT_2016}. 
In higher dimensions, e.g., in two dimension, the model with Dirac fermions in the presence of random vector potential also hosts such critically delocalized states~\cite{Motrunich02}. 
%
%

In recent years, the concept of single-particle Anderson transition has been generalized to the many-body case, generating the field now known as many-body localization~(MBL)~\cite{Basko06, gornyi2005interacting, nandkishore2015many, abanin2017recent,ALET2018498,CollAba}. 
The interplay between disorder and electron-electron interactions induce a quantum phase transition, separating a thermal~(ergodic) phase from a many-body localized one even at high energy density~\cite{Basko06, gornyi2005interacting, Pal10, oganesyan2007localization, Bera15}.
The  MBL transition is known to be an eigenstate quantum phase transition~\cite{huse2015review}, occurring at the level of single-eigenstates. 
For instance, the bipartite entanglement entropy calculated in an eigenstate changes from being volume law in the ergodic phase to an area law in the localized one. 

By now several indicators are known to characterize the MBL phase, for instance, the absence of thermalization, Poisson level spacing distribution and exponential decay of correlation functions are few of those. 
Importantly, the absence of thermalization in the MBL phase is induced by an emergent form of integrability~\cite{imbrie2014many,huse2013phenomenology,rms-IOM,  Mukerjee2016integrals,Chandran2014, rademaker2015, monthus2015integrals, serbyn2013local, GDT_ent_19}, described by the existence of an extensive number of quasi-local integrals of motion. 
The presence of interactions induces dephasing between the local integrals of motion, giving rise to a slow propagation of information through the entire system even though charge or energy transport is prohibited by localization~\cite{bardarson2012unbounded,Pro08,serbyn2013universal,GDT_ent_19}. However, a complete understanding of the MBL phase and its associated critical properties are still lacking due to significant finite size and finite time effects in numerical simulations~\cite{Bera_slow_19, PandaMBL19, Abanin_2019, Prosen2019no_MBL, Bera17}.
%

%
%
%

As in Anderson localization, the presence of symmetries also gives rise to richer physics in the MBL transition~\cite{Vass16,VassPotter16, ALET2018498, Protopopov17, Chan18, Wahl_top_18,Friedman_Aaron18,Gopalakrishnan19,Protopopov_Sca_19, Brenes18, huse2013localization, Kill14}. 
Recently, it has been shown that the presence of continuous symmetries such as the SU(2) one,  alters the eigenstates properties of the MBL phase. 
For instance, the bipartite entanglement entropy in highly excited eigenstates shows logarithmic scaling with the system size instead of a conventional area law~\cite{Protopopov17, YuPRB18, Protopopov_Sca_19}. 
The sub-thermal scaling of the entanglement is attributed to the resonance structure of the eigenstates that originates from the symmetry of the Hamiltonian.
As a result, an SU(2) symmetric model shows unconventional spin or charge transport as compared to the usual MBL phase~\cite{SpinChargeprbDelande18, Protopopov17, Kozarzewski19, Protopopov19}.  
In the spinfull Hubbard model this is manifested in the subdiffusive propagation of spin excitations in the background of localized charge degrees of freedom.
It is believed that eventually the spin excitations will act as a bath for the localized charge and will delocalize it in the thermodynamic limit~\cite{Protopopov19, Sroda19}. 
%

A different type of symmetry is the chiral or particle-hole~(PH) symmetry, which is the main topic of this work, also known to give rise to a distinct character of the MBL phase~\cite{Kill14,Vass16}. 
In the presence of PH symmetry the non-interacting ground state is described by random singlets~\cite{Fisher94}. 
Using real space renormalization group calculations, it has been shown that the random singlet phase also extends to excited states, and is dubbed as quantum critical glass~\cite{Vass16}. 
On the contrary, in the presence of both interactions and PH symmetry, the highly excited states spontaneously break the original symmetry at strong disorder, therefore giving rise to an MBL phase. 
However, the  MBL phase with PH-symmetry differs from the well known MBL phase by the fact that the excited states are macroscopic cat states, which are protected by the global $\mathbb{Z}_2$ symmetry. 
The spectral pairing of eigenstates is responsible for a non-zero value of the Anderson-Edwards order parameter, which further implies that the PH symmetric MBL phase is of spin-glass nature~\cite{Vass16, Kill14}.

Several eigenstate properties of the PH symmetric MBL phase have been explored, but the dynamical properties have so far not been extensively investigated~\cite{Vass16, Kill14}.  
In this work, we explore the dynamics of an MBL phase in the presence of PH symmetry. 
We focus on the time evolved entanglement entropy quenching from a typical product state to probe information propagation, along with the infinite temperature density-density correlator for particle transport. 
We observe that in the PH symmetric MBL phase the entanglement entropy grows logarithmic-like at weak interactions and algebraically for stronger, in contrast to usual MBL where it grows logarithmically at all interaction strengths~\cite{bardarson2012unbounded,Pro08,serbyn2013universal,GDT_ent_19, GDT_frag_2019}.
Moreover, at asymptotically large times in a finite system the entanglement entropy saturates to a non-ergodic extensive value.
The density-density correlation shows propagation in space and time pointing towards an unconventional non-ergodic phase, which might eventually delocalize at very long times in the thermodynamic limit.
We qualitatively explain our findings by introducing an effective model for the dynamics in the two-particle sector of the original model.
We observe that even at this zero-density limit the essential features of the finite-density dynamics are revealed via bound states propagation of the two particles due to the interaction.
We point out that such a propagation arises due to long-range resonances protected by the underlying PH symmetry.

The  rest  of  the work is  organized  as  follows. In Sec.~\ref{sec:model}, we introduce the interacting PH symmetric model and we discuss its non-interacting limit. 
Here, we also define the observables that we consider in this work, e.g., entanglement entropy, and infinite temperature density-density correlator.
In Sec.~\ref{sec:Numerics}, we present the numerical results. 
In particular, Sec.~\ref{sec:Numerics1} shows the spectral and eigenstates properties of the model that also confirm the existence of an MBL phase at strong disorder. 
While in Sec.~\ref{sec:Numerics2}, we focus on the out-of-equilibrium dynamics at strong disorder. 
In Sec.~\ref{sec:Two_par}, we provide evidence that the same type of dynamics is also present in the case where we consider only two particles in the system.

\section{Model and Methods}\label{sec:model}
In order to study the dynamical properties of a strongly disordered MBL phase with PH symmetry, we use the following $t{-}V$ spinless fermionic model with bond disorder and next nearest neighbor interactions,
\eq{
\msr{H} = \sum_{i=-L/2}^{L/2-2} J_i \Le(c^\dg_i c_{i+1} + \text{h.c.}\Ri) + V\Le(n_i-\f{1}{2}\Ri) \Le(n_{i+1}-\f{1}{2}\Ri),
\label{eq:ham}
}
where $c_i^\dg~(c_i)$ is the fermionic creation~(destruction) operator at site $i$ and $n_i=c_i^\dg c_i$ is the local density operator. $L$ is the system size and $N=L/2$ is the number of fermions.
The $J_i$'s are the random bond disorder defined by $J_i = e^{\mu_i}$, where $\mu_i$'s are independent random variables uniformly distributed between $[-W, W]$. 
This choice ensures that $J_i$'s are positive. $V$ is the strength of the interaction. 
The model in Eq.~\ref{eq:ham} is equivalent, via a Jordan-Wigner transformation to a spin-$1/2$ XXZ Heisenberg chain with random bonds.
Moreover, $\msr{H}$ has a global $\mathbb{Z}_2$ symmetry~(PH symmetry) generated by the parity operator $P = \prod_i \sigma_i^x$ ($P c_i P^{-1} = (-1)^{i+1} c_i^\dagger$).
%
%

The non-interacting case ($V=0$) is known as the Dyson model ($XX$ model with random bonds) and is integrable~\cite{Dyson53,Weissmann_1975, Theo76, Eggarter78, Fleishman_1977,Souk81,Brower2000, Bush_1975, Ziman82}.
In this limit the single-particle density of states shows a divergence at zero energy $\varrho(E) \sim 1/ (E |\log^3(E)|)$ as $E\rightarrow 0$~\cite{Dyson53,Brower2000, Ziman82}.
The presence of this divergence implies through the  Herbert-Jones-Thouless relation~\footnote{$\xi_{\text{loc}}^{-1}(E) = \int dE' \rho(E') \log{|E-E'|}$.}, a logarithmic divergence for the localization length $\xi_{\text{loc}}$ with energy~\cite{Ziman82, Theo76}. 
Nevertheless, the zero energy states are not extended and they can be considered quasi-localized (marginally localized), meaning that the localization length depends only sub-extensively  on system size $\xi_{\text{loc}} \sim \sqrt{L}$. Instead, the eigenstates away for the divergence, are exponentially localized $\xi_{\text{loc}} \sim \mathcal{O}(L^0)$~\cite{Ziman82, Theo76, GDT_2016}, similar to the case with diagonal disorder. 
Consequently, the non-equilibrium dynamics of the non-interacting model show a sub-logarithmically slow entanglement propagation $\sim \log(\log(t))$~\cite{Chiara_2006, Vosk_2013, Sirker_16, GDT_2016} and charge transport~\cite{Krapivsky_2011, GDT_2016}.

In Ref.~\onlinecite{Vass16} a similar model of Eq.~\ref{eq:ham} has been considered, where using a combination of renormalization group method and exact diagonalization the existence of a strong disorderd MBL spinglass phase has been established. 
With the aim to detect such a putative MBL phase for the considered model (Eq.~\ref{eq:ham}), we compute the following two conventional indicators. 
First, the level spacing statistics in the middle of the spectrum.
The level spacing statistics is defined as $r_n=\text{min}(\delta_n, \delta_{n+1})/\text{max}(\delta_n, \delta_{n+1})$, where $\delta_n=E_n-E_{n-1}$ with $E_n$'s are the ordered eigenenergies of $\msr{H}$.  
Second, we compute the disordered averaged bipartite entanglement entropy $\mathcal{S}(A)=-\text{Tr} \rho_{L/2} \log(\rho_{L/2})$, where $\rho_{L/2}$ is the reduced density matrix of half of the system in an eigenstate chosen from the middle of the spectrum. 
The eigenvalue and eigenstate properties are calculated using standard shift-invert diagonalization techniques~\footnote{We computed $32$ eigenstates and eigenvalues for any random configurations.}.

After validating the existence of a non-ergodic phase in this model, we probe the dynamics of such phase via the propagation of the bipartite entanglement following a quantum quench and the density-density correlator at infinite temperature, defined as,
\eq{
\label{eq:correlation}
 \msr{C}(i,t)= \f{1}{\msr{D}}\overline{\text{Tr} \Le( \hat n_i(t)-\frac{1}{2}\Ri) \Le(\hat n_{0}-\frac{1}{2} \Ri) } \Theta(t), 
}
where $\text{Tr}$ represents the infinite temperature average and $\msr{D}$~\footnote{$\msr{D}= \binom{L}{N}$.} is the dimension of the many-body Hilbert space. 
We renormalize  the  correlator $\msr{C}(i,t)$  via  its  Fourier transform $\msr{C}(q,t) =\mathcal{F}_q[\msr{C}(i,t)]$~\eq{\label{eq:correlation1}
\Pi(i,t) = \mathcal{F}_i \left [ \frac{\msr{C}(q,t) }{\msr{C}(q,t=0^+)} \right ].
}
The density-correlator $\Pi(i,t)$ can be interpreted as a probability distribution $\Pi(i,t)\ge 0$, $\sum_i \Pi(i,t)=1$ and $\Pi(i,t=0^+) = \delta_{i0}$.

To monitor the dynamics of the system we focus on the second moment of the correlator in Eq.~\ref{eq:correlation1}, 
\eq{
\Delta x(t)= \left[ \sum_{i=-L/2}^{L/2} i^2 \Pi(i,t)  - (\sum_{i=-L/2}^{L/2} i \Pi(i,t) )^2\right]^{1/2}. 
\label{eq:dex}
}
To access larger system sizes, here we use Chebyshev polynomials techniques for the time evolution and evaluate $\text{Tr}$ stochastically using random states~\cite{Wei06, Bera17}.

In a conventional localized phase, $\Delta x(t)$ probes the localization properties of the system~\cite{Pro08, Bera17}, saturating at a finite $L{-}$independent  value in the limit of asymptomatically long times. Indeed, its saturation value $\Delta x(t {\rightarrow} \infty)$ could be used to define a localization length in analogy with the non-interacting model. 
In an MBL phase without special symmetries a saturation of $\Delta x(t)$ is expected at long-times, however, recently it has been shown that due to some residual dynamics the time scale to reach such a saturation could be much longer compared to the non-interacting limit~\cite{Bera_slow_19}. 

\section{Numerical results}\label{sec:Numerics}
\begin{figure}[tb]
\centering
 \includegraphics[width=1\columnwidth,keepaspectratio=true]{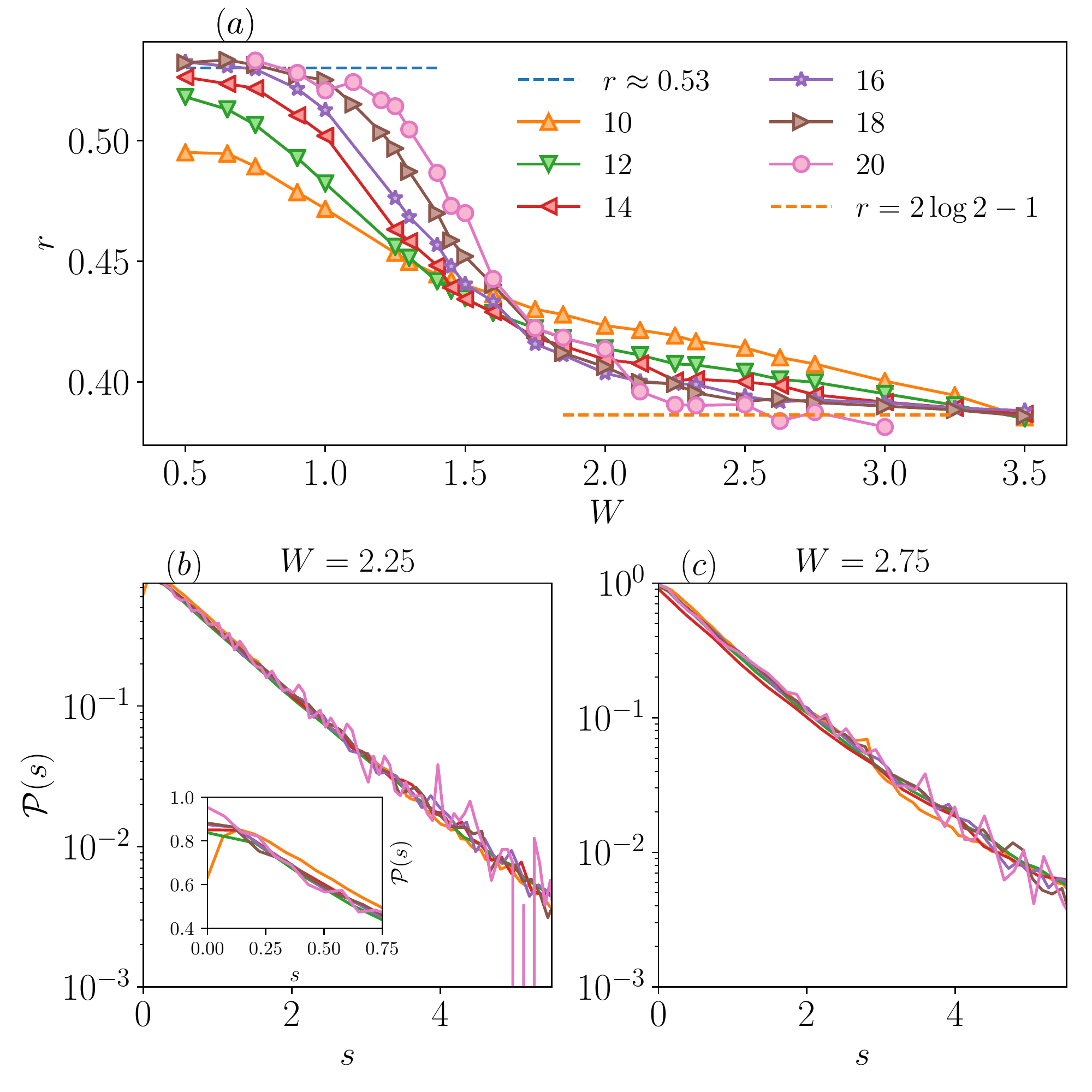}
\caption{(a): Level spacing parameter $r$ as a function of disorder strength $W$ for different system sizes $L$ and $V=2$. The dashed lines are the Poissonian value $r=2\log{2}-1$ and the GOE one $r\approx 0.53$. (b)-(c): Probability distribution $\mathcal{P}(s)$ for the renomalized level spacing $s = E_{n+1}-E_{n}/\la  E_{n+1}-E_{n} \ra$ for two values of disorder strength $W=2.25,2.75$, respectively. The inset in (b) shows a magnification of $\mathcal{P}(s)$ for small values of $s$.} 
\label{f1_r}
\end{figure}
%
\subsection{Spectral and eigenstates properties}\label{sec:Numerics1}
In this section, we inspect spectral and eigenstates properties in the middle of the spectrum of $\msr{H}$ in Eq.~\ref{eq:ham} with the aim to establish the existence of an MBL phase at strong disorder. Moreover, we restrict the Hamiltonian to its positive $\mathbb{Z}_2$ sector~\cite{Vass16}.

{\bf Level spacing statistics}:
In an MBL phase due to an emergent form of integrability, energy levels tend to cross each other and the  probability distribution for the level spacing is expected to be Poissonian. 
As a consequence, the level spacing statistics, $r$ takes the value $2\log{2} - 1$~\cite{Atas_2013, oganesyan2007localization}. Instead, in an ergodic phase of $\msr{H}$ energy levels avoid each other and $ r $ takes the same value of a random matrix as known for the Gaussian Orthogonal Ensemble (GOE) $ r \approx 0.53$~\cite{Atas_2013, oganesyan2007localization}.

Figure~\ref{f1_r}~(a) shows the level spacing statistics, $ r $, as a function of disorder $W$ and fix interactions strength $V=2$.  
With increasing disorder strength the $ r $ changes from GOE $\simeq 0.53$ to Poisson statistics $\simeq 2\log{2}-1 \approx 0.386$. 
Approximately the transition happens at disorder strength $W_{\text{c}} \lesssim 2$. 
However, strong finite size effects are manifested by a shifting of $W_c$ with $L$ through ergodicity.
This aforementioned shift has been seen also for the model with diagonal random disorder~\cite{alet2015, Kill14, Prosen2019no_MBL, Abanin_2019, SierantThoulessTime19, LevelStatJakubPRB19}, however, it seems to be more pronounced here. 
Due to unavailability of larger system sizes and shifting of the crossing point of the $r$ we do not attempt to do a finite-size collapse of the data. 
However, to better understand the stability of the localized phase, we inspect the probability distribution $\mathcal{P}(s)$ of the energy level spacing $s = E_{n+1}-E_{n}/ \la E_{n+1}-E_{n} \ra$~\footnote{$\int s \mathcal{P}(s) ds=1$}. 
Figures~\ref{f1_r}~(b)-(c) show $\mathcal{P}(s)$ for several system sizes $L=\{10-20\}$ at strong disorder  $W=2.25, 2.75$ respectively, which clearly support the expected Poissonian statistics (see inset in Fig.~\ref{f1_r}~(b)).
%

\begin{figure}[tbh]
\centering
 \includegraphics[width=1\columnwidth,keepaspectratio=true]{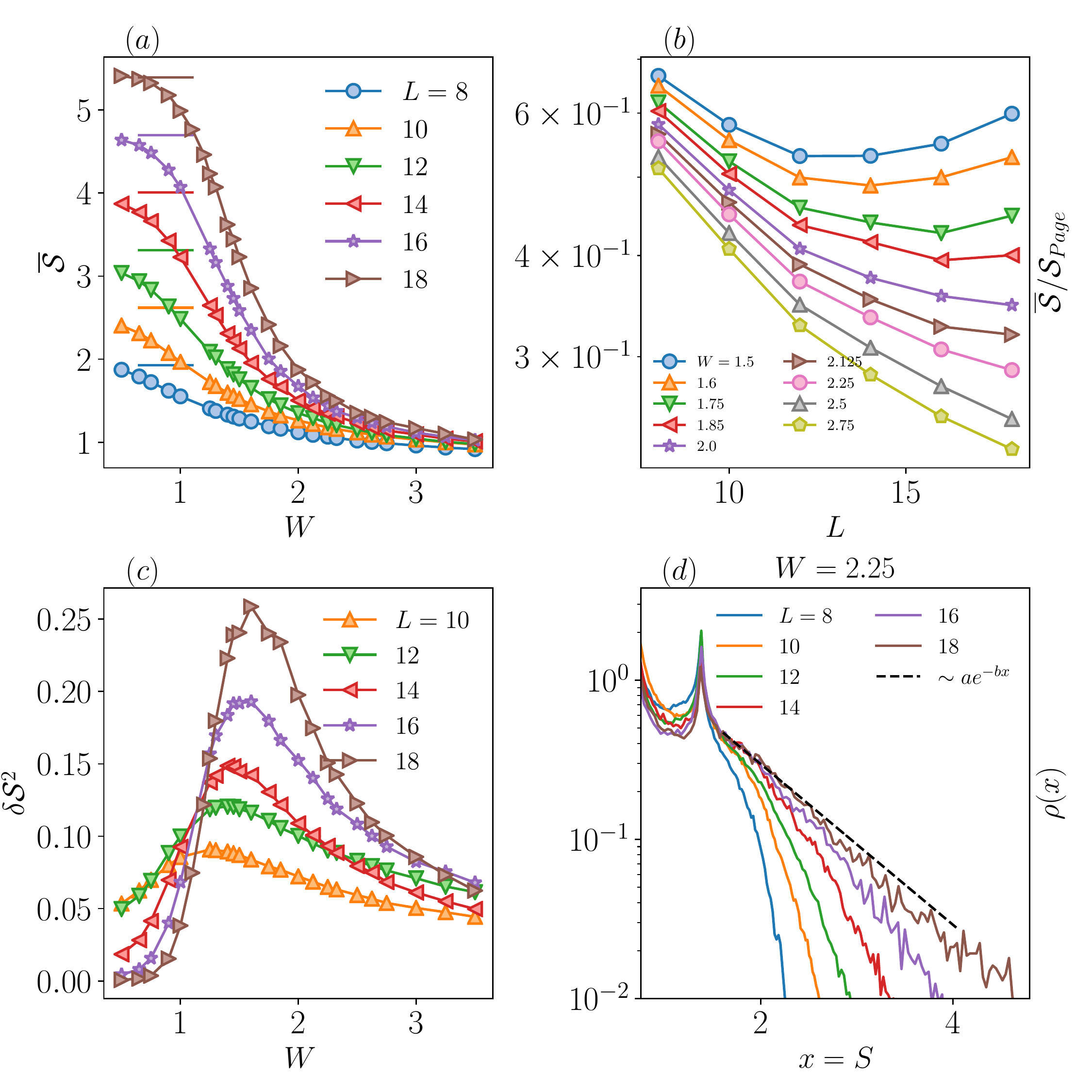}
\caption{(a): Averaged half-chain entanglement entropy $\overline{\mathcal{S}}$ as a function of disorder strength $W$, several $L$ and $V=2$. The dashed lines are the entanglement entropy of a random state $\mathcal{S}_{Page} = ((L-1)\log{2}-1)/2)$. (b): $\overline{\mathcal{S}}/\mathcal{S}_{Page}$ as function of $L$ and several $W$ close to the putative MBL transition. (c): Variance $\overline{\delta \mathcal{S}^2}$ of the entanglement entropy as function of $W$. (d): Probability distribution $\rho(x)$ of the entanglement entropy $x=\mathcal{S}$ for several $L$ at $W=2.25$. The dashed line is a guide for eye to support the
exponential form $\rho(x\gg1) \sim a e^{-b x}$ ($a\approx 3.1$, $b\approx 1.2$).}
\label{f2_entg}
\end{figure}
{\bf Bipartite entanglement entropy}: 
The entanglement entropy $\mathcal{S}$ is a useful tool to probe ergodicity and its break-down. 
In an MBL phase, eigenstates are only locally entangled, implying an area law scaling for the entanglement entropy $\mathcal{S}\sim \mathcal{O}(L^0)$. 
In contrast, in a thermal phase eigenstates are highly entangled and $\mathcal{S}$ scales linearly with $L$ (volume law).

Figure~\ref{f2_entg}~(a) shows the disorder averaged eigenstate half-system entanglement $\mathcal{\mathcal{S}}$ as a function of disorder strength $W$. 
At weak disorder $W \lesssim 2.0$ in the ergodic phase one clearly observes the expected volume law, in agreement with the GOE value of the level spacing parameter $r$.
Moreover, $\overline{\mathcal{S}}$ approaches the value $\mathcal{S}_{Page}= ((L-1) \log{2}- 1)/2$ of a random state in the Hilbert space~\cite{Page1993}, which is the expected behavior for $\mathcal{S}$ for thermal states at infinite temperature~\footnote{Here at infinite temperature refers to the middle of the energy spectrum}.
For larger disorder, within the available system sizes, we also observe a weak flow of entanglement with system size, although it is far from a volume law.
In Fig.~\ref{f2_entg}~(b) we summarize this weak flow by plotting $\overline{\mathcal{S}}/\mathcal{S}_{Page}$ as a function of system sizes for different values of disorder. 
For weak disorder $W\approx 1.5 - 1.85$, but close to the transition, we report an upturn of $\mathcal{S}/\mathcal{S}_{Page}$ with increasing system sizes. 
At this disorder value the system shows an area law scaling as it goes down linearly with $L$.
For larger $L \gtrsim 12$ the system is out of the `correlation volume' and eventually returns to volume law scaling as one would expect. 
The correlation volume increases quite rapidly with increasing disorder as can be see in Fig.~\ref{f2_entg}~(b). 
For stronger disorder $W\approx 2.0$, $\mathcal{S}/\mathcal{S}_{Page}$ decreases and no minimum is reached within the available system sizes. 
What happens with increasing system sizes and whether at larger disorder values one observes volume law scaling of entanglement (curves show the upturn) cannot be ruled out from the existing data. 
This phenomena is reminiscent of the situation in the Anderson model on the random regular graph~(RRG)~\cite{ Tik16, Tikh2019_K(w), Deluca14, Kra18, Alts16, bera19, DGT_sub_19}. 
In RRG it has also been shown that the correlation volume is exponentially large in disorder strength, therefore finite-size numerics underestimate the true critical point of the transition~\cite{Tik16, Kra18, Alts16}. 
Figure~\ref{f2_entg}~(c) shows the sample variance of the bipartite entanglement entropy $\delta S^2$~\footnote{$\overline{\delta S^2} = \overline{S^2} -\overline{S}^2$  and the average is taken over both disorder and eigenstates.} as a function of disorder strength for several system sizes.
The variance is growing with $L$ close to the estimated finite-size transition as also observed in diagonal disorder model~\cite{alet2015, Kill14, GDT_frag_2019}.  

Like for the level spacing parameter $ r $ to better understand the nature of the MBL phase, we study the probability distribution $\rho(\mathcal{S})$ of the bipartite entanglement entropy $\mathcal{S}$, as shown in Fig.~\ref{f2_entg}~(d).
As one can notice the shape of $\rho(\mathcal{S})$ is highly non-thermal. Moreover, the distribution function shows a clear peak structure at $\mathcal{S}\approx \log(2)$. 
The origin of the peak can be understood by recalling that the PH symmetric MBL phase is of spin-glass nature. 
It implies that the eigenstates are macroscopic cat states $|\Psi_\text{c} \ra =  (|E_n\ra \pm P|E_n\ra)/\sqrt{2}$, where $P|E_n\ra$ denotes the symmetry reversed state of $|E_n\ra$.
These type of states contribute to the entanglement entropy as $\mathcal{S}\approx \log(2)$ as seen with a pronounced peak in the entanglement distribution. Importantly, $\rho(\mathcal{S})$ develops exponential tails $(\rho(\mathcal{S})\sim e^{-b \mathcal{S}})$, what might imply an area law scaling for $\overline{\mathcal{S}}$ ($\sim \mathcal{O}(L^0)$) in the thermodynamic limit. Nevertheless, logarithmic corrections cannot be ruled out, e.g. $\overline{\mathcal{S}} \sim \log{L}$.

From these observables here we conclude that indeed there exists an ergodic-MBL transition in finite size for this PH symmetric interacting model. In the following sections we discuss the dynamical properties of this MBL phase. 

\begin{figure}[tb]
\centering
 \includegraphics[width=0.9\columnwidth,keepaspectratio=true]{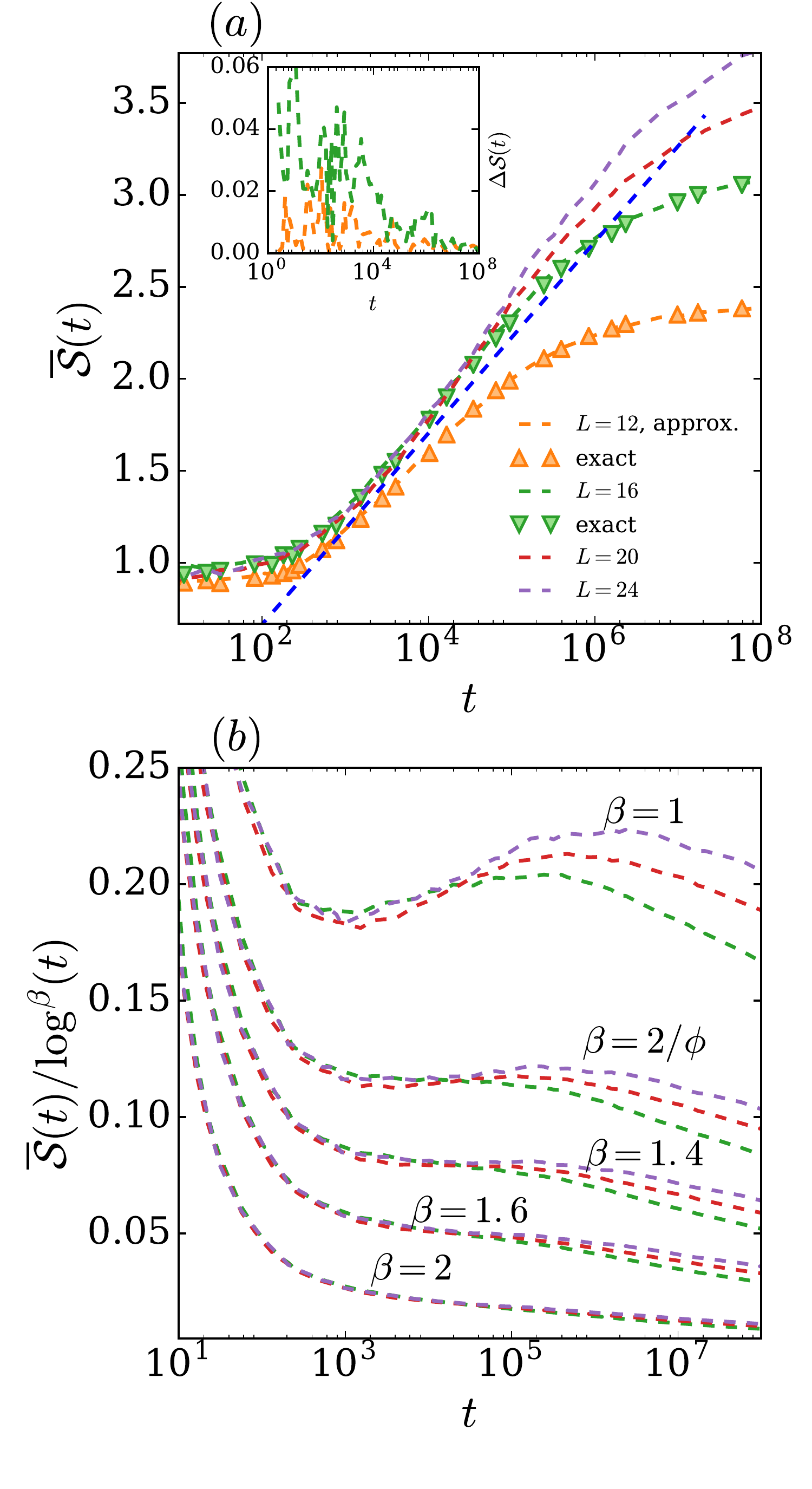}
\caption{(a): The averaged bipartite entanglement entropy $\overline{\mathcal{S}}(t)$ at strong disorder $W=2.5$ and weak interactions $V=0.01$. The dots curves (exact) are obtained preforming the time evolution with the full Hamiltonian $\mathcal{H}$ in Eq.~\ref{eq:ham}, while the dashed ones (approx.) with the effective Hamiltonian in Eq.~\ref{eq:H_eff_ent}. The blue straight dashed line is just for guide of eyes $\sim \log (t)$. The inset shows the relative fluctuations $\Delta \mathcal{S}(t) = \overline{|\mathcal{S}(t) - \mathcal{S}^{\text{Approx.}}(t)|}/\overline{\mathcal{S}}(t)$ for $L\in \{12, 16\}$.
(b): $\overline{\mathcal{S}}/\log^\beta(t)$ for several $L\in \{16, 20,24\}$ and several $\beta \in [1,2]$.  }
\label{f3_entg_dyn_weak}
\end{figure}

\subsection{Out-of-equilibrium dynamics}\label{sec:Numerics2}
Having established the existence of a non-ergodic phase, we now turn to probe its dynamical properties. By studying the dynamics of the system, we will be able to make definite statement about the nature of the MBL phase for finite time scales. 
%
For instance, we find that the dynamics is vastly different from the conventional MBL phase~\cite{Pro08,Reich15, Bera_slow_19, Bera17, GDT_frag_2019, La16, GDT_ent_19} even though the system shows Poissonian level statistics. 

\begin{figure}[tb]
\centering
 \includegraphics[width=1.\columnwidth,keepaspectratio=true]{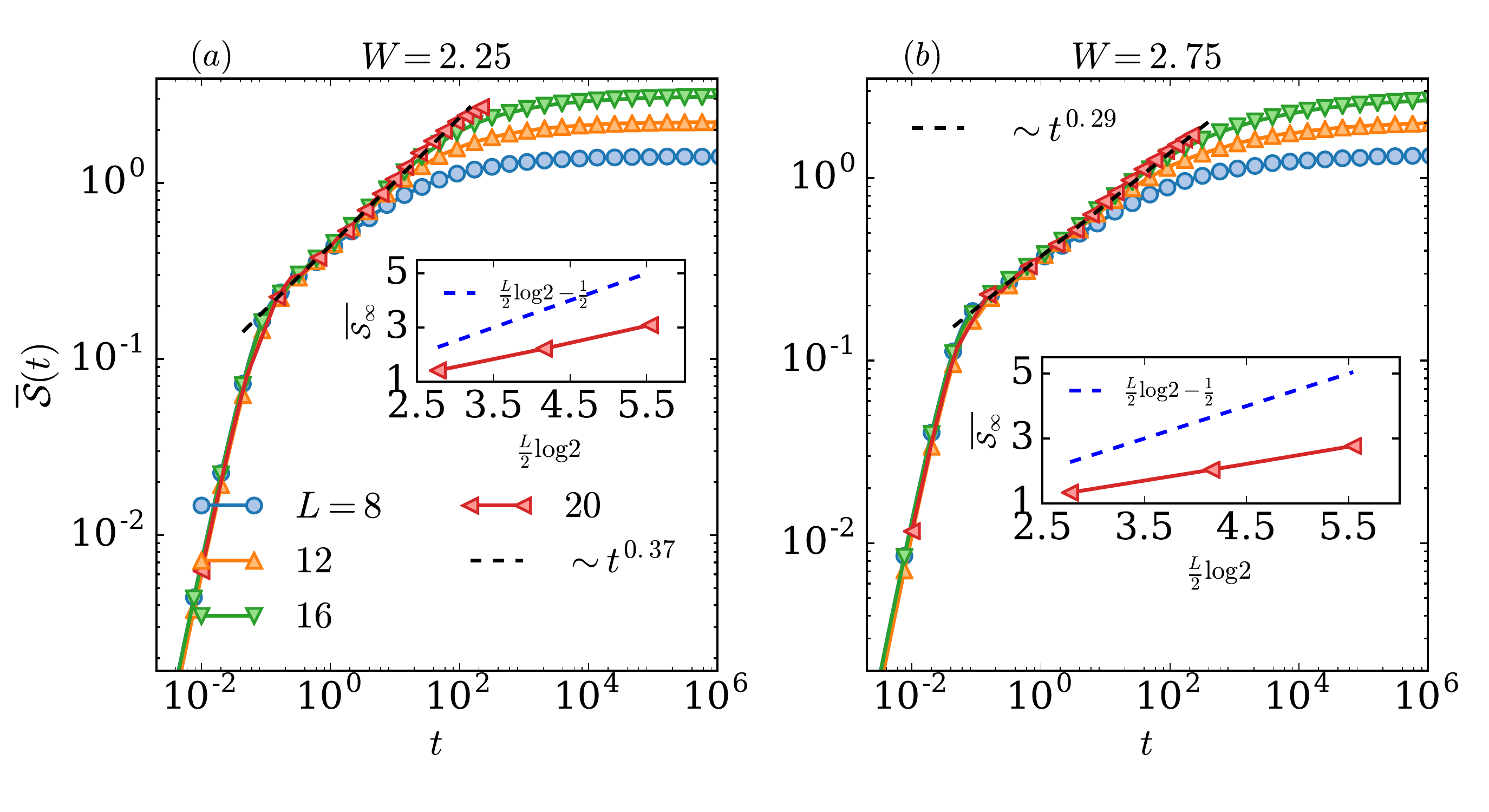}
\caption{(a)-(b): Entanglement growth $\overline{\mathcal{S}}(t)$ quenching a typical product state for two values of $W = 2.25, 2.75$ and $V=2$. The dashed line are a possible fits, in order to show the algebraic growth propagation $\overline{\mathcal{S}}(t)\sim t^{\gamma}$. }
\label{f3_entg_dyn}
\end{figure}
\begin{figure}[tb]
\centering
 \includegraphics[width=1\columnwidth,keepaspectratio=true]{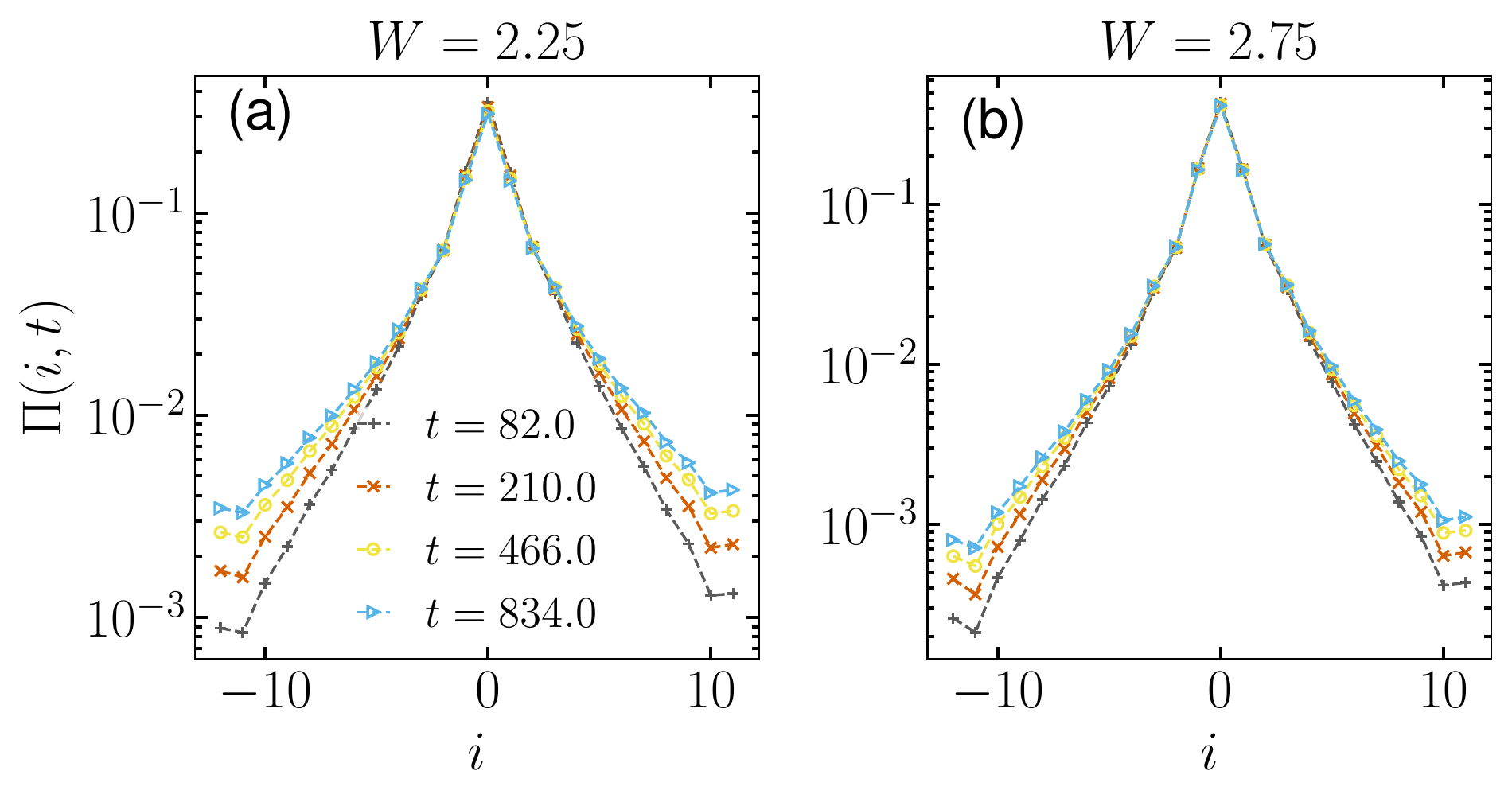}
\caption{$\Pi(i,t)$ in the localized phase for $L=24$ and $V=2.0$. As observed, the density correlator predominantly shows exponential tails, however, with increasing time the tail is lifted, which facilitate the charge propagation.}
\label{f4_distr}
\end{figure}
{\bf Entanglement growth}:
First, we probe the propagation of entanglement by quenching random product states of the form $\prod_s c^\dagger_{i_s}|0\ra$. 
In the usual exponentially localized MBL phase, in which the quasi-local integrals of motion develop exponential tails, the entanglement grows logarithmically in time ($\mathcal{S}(t)\sim \log(t)$) due to an interaction-induced dephasing mechanism~\cite{bardarson2012unbounded,Pro08,serbyn2013universal,GDT_ent_19}. 

Let us start by investigating the entanglement propagation in the presence of weak interactions. 
For accessing larger system sizes at arbitrary large time scales, we use a recently proposed method, which is able to quantify the induced dephasing mechanism at strong disorder and weak interactions~\cite{GDT_ent_19}. 
This perturbative method has been already used to study information propagation in one- and two-dimensional MBL systems~\cite{GDT_ent_19}, in an algebraic MBL phase~\cite{GDT_alg_19}, and to inspect transport properties of MBL systems weakly coupled to a thermal bath~\cite{Wu_2019}.
The method relies on the perturbative nature of the integrals of motion.
As a consequence, in the limit of weak interactions the integrals of motion of $\mathcal{H}$ in Eq.~\ref{eq:ham} can be taken the ones of the non-interacting case.

Rewriting the Hamiltonian in terms of the  creation  (annihilation) operator $\eta^\dagger_k$ ($\eta_k$) for the single-particle eigenstate $\psi_k(x)$ with eigenvalue $\epsilon_k$, we obtain
\eq{ \msr{H} = \sum_k \epsilon_k \eta_k^{\dagger} \eta_k + V\sum_{l,n,p,q} B_{l,n,p,q} \eta_l^\dagger \eta_n \eta_p^\dagger \eta_q,
}
where $B_{l,n,p,q} = \sum_{i} \psi_l(i) \psi_n(i)\psi_p(i+1)\psi_q(i+1)$~\cite{GDT_ent_19}.
Discarding the terms that do not commute with the non-interacting integrals of motion $\tau_k = \eta_k^\dagger \eta_k$ ($[\mathcal{H}(V=0), \tau_k]=0$), we obtain the following effective Hamiltonian 
\begin{equation}
\label{eq:H_eff_ent}
 \msr{H}^{\text{eff}} = \sum_k \epsilon_k \tau_k + V\sum_{l,n} S_{l,n} \tau_l \tau_n,
\end{equation}
where $S_{l,n} = B_{l,l,n,n} - B_{l,n,l,n}$.  

We benchmark these approximations by comparing the propagation of $\mathcal{S}(t)$ generated with the exact dynamics of $\mathcal{H}$ in Eq.~\ref{eq:ham} and with  $\mathcal{H}^{\text{eff}}$ in Eq.~\ref{eq:H_eff_ent}. 
Figure~\ref{f3_entg_dyn_weak} shows the averaged bipartite entanglement entropy $\overline{\mathcal{S}}(t)$ at strong disorder $W=2.5$ and weak interactions $V=0.01$ computed with the full Hamiltonian $\mathcal{H}$ (dots, exact) and $\mathcal{H}^{\text{eff}}$ (dashed line, approx.) for $L\le 16$.  The curves generated with $\mathcal{H}$ and $\mathcal{H}^{\text{eff}}$ are almost indistinguishable. Moreover, to better quantify the difference, we compute the relative error 
$\Delta \mathcal{S}(t) = \overline{|\mathcal{S}(t) - \mathcal{S}^{\text{Approx.}}(t)|}/\overline{\mathcal{S}}(t)$ as function of time for system sizes reachable with exact diagonalization (see the inset of Fig.~\ref{f3_entg_dyn_weak}). $\Delta \mathcal{S}(t)$ provides numerical evidence that the method is reliable in the considered regime, since it is bounded with time, does not present strong finite-size effects with $L$ and the deviation is smaller than $6\%$.

Thus after having validated the robustness of the method for strong disorder and weak interactions, we now compute $\overline{\mathcal{S}}(t)$ performing the dynamics with $\mathcal{H}^{\text{eff}}$ in Eq.~\ref{eq:H_eff_ent}. 
At time scales  governed  by  the  interaction strength  $t^\star \sim 1/V\approx 100$ a slow (presumably logarithmic) propagation takes place, as shown in Fig.~\ref{f3_entg_dyn_weak}~(a) for system sizes up to $L\le 24$.
We detect a small positive curvature in log-scale for the growth of $\overline{S}(t)$, implying that the propagation might be described  by $\overline{S}(t)\sim \log^\beta (t)$ with $\beta >1$. 
With the aim to better understand the behavior of $\overline{S}(t)$, we analyze the function $\overline{S}(t)/\log^\beta(t)$ by tuning the parameter $\beta$. 
Figure~\ref{f3_entg_dyn_weak}~(b) shows $\overline{S}(t)/\log^\beta(t)$ as function of time with $\beta \in [1,2]$ and several $L$. For $\beta = 1$, $\overline{S}(t)/\log^\beta(t)$ increases with time, while for $\beta =2$ decreases. As consequence $\overline{S}(t) \sim \log^\beta(t)$ with $1<\beta <2$.
We found the best fit for $\beta \approx 1.4$, for which  $\overline{S}(t)/\log^\beta(t)$ develops a plateau with respect to $t$. 

The slow propagation at strong disorder and weak interactions $\overline{S}(t) \sim \log^\beta(t)$ with $\beta \approx 1.4$ is consistent with the renormalization group calculation proposed in Ref.~\onlinecite{Vosk_2013}, $\overline{S}(t) \sim \log^{2/\phi}(t)$ with $\phi = (1+\sqrt{5})/2$ being the golden ratio. 
It is important to underline that this result could not be found by inspecting small sizes $L\le 16$, due to the strong finite effects even at short times, i.e., the dynamics for $L=12$ and $L=16$ overlap only at small transient times (see Fig.~\ref{f3_entg_dyn_weak}~(a)). Moreover, the applied method could be seen as an indirect proof, of the fact that at weak interactions the entanglement growth is caused by the interaction-induced dephasing mechanism (Eq.~\ref{eq:H_eff_ent}), at least for the inspected time scales $t \approx 10^{6}$.

Next, we study information propagation for stronger interaction strengths $V$, upon keeping the disorder $W$ strong, in order to remain in the MBL phase of $\mathcal{H}$. In this regime the  pertubative argument (Eq.~\ref{eq:H_eff_ent}) might break down and the off-diagonal terms $\eta^\dagger_l \eta_n \eta^\dagger_p \eta_q$ would play an important role in the information propagation. Due to non-applicability of the aforementioned method, we compute $\mathcal{S}(t)$ using exact diagonalization for $L\le 16$ and Chebyshev integration techniques for $L=20$. 

In Figs.~\ref{f3_entg_dyn}~(a)-(b) the growth of the averaged entanglement $\overline{\mathcal{S}}(t)$ is shown for system sizes $L=8-20$ in a double logarithmic plot for values of $W=2.25, 2.75$ and $V=2$.
It is clearly seen that in the spin-glass MBL phase~($W=2.25, 2.75$) the growth of entanglement is faster than in an exponential MBL phase and it is consistent with an algebraic behavior $(\overline{\mathcal{S}}(t) \sim t^{\gamma})$. 
It is important to point out that although the entanglement propagation is rather fast, even then the system does not reach a thermal value in the limit of asymptotic large times for finite system. 
In the insets of Figs.~\ref{f3_entg_dyn}~(a)-(b), we show the long-time saturation value $\overline{\mathcal{S}}_\infty = \lim_{T\rightarrow \infty} 1/T\int_0^\infty \overline{\mathcal{S}} (t) dt$ of $\overline{\mathcal{S}}(t)$. $\overline{\mathcal{S}_\infty}$ follows a volume law $(\overline{\mathcal{S}_\infty} \sim L)$ though the value itself is non-thermal at infinite temperature, i.e., $\overline{\mathcal{S}_\infty}/\Le(L/2 \log{2}\Ri) <1$~\cite{bardarson2012unbounded}.

\begin{figure}[tb]
\centering
\includegraphics[width=1\columnwidth,keepaspectratio=true]{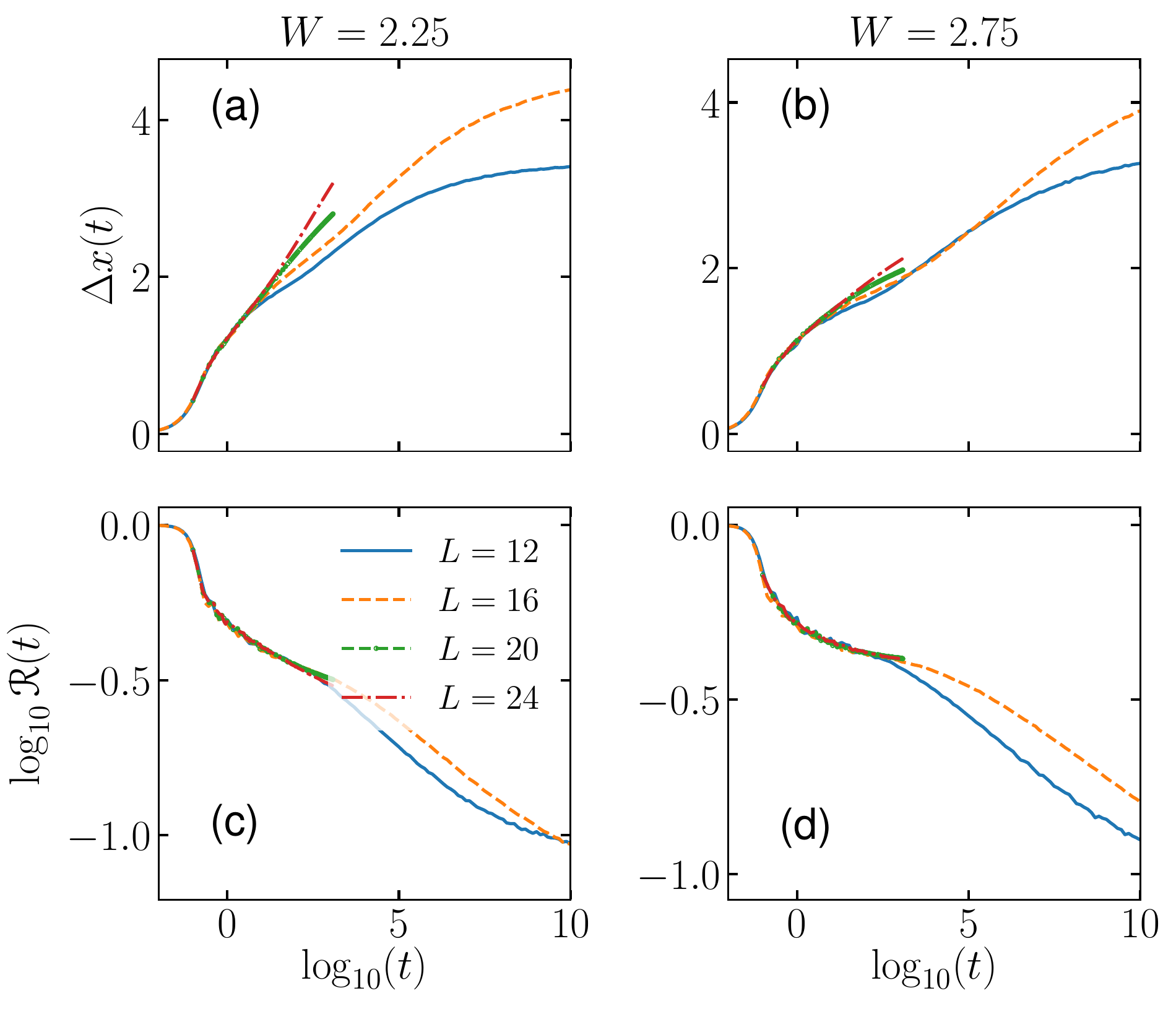}
\caption{(a)-(b): $\Delta x(t)$ in the localized phase for different $L$, $V=2.0$ and two values of $W = 2.25, 2.75$, respectively. (c)-(d) Return probability $\mathcal{R}(t) = \Pi(i=0,t)$ for the same values of $L$, $V$ and $W$.} 
\label{f5_dex}
\end{figure}
{\bf Density-density correlator}: Next, we study the charge relaxation in the system through the density-density correlator $\Pi(i,t)$ defined in Eq.~\ref{eq:correlation}. 

Figures~\ref{f4_distr}~(a)-(b) show  $\Pi(i, t)$  for system size $L=24$, and for two disorder values in the localized phase $W=2.25, 2.75$ and $V=2$. The probability distribution function $\Pi(i,t)$ is shown at different times to emphasize that even at the longest observation time it is not stationary. $\Pi(i,t)$ develops tails, which moves upwards with increasing time. While the core of the distribution seems to be more stable. To capture the movement of the tail, we further calculate the second moment of $\Pi(i,t)$ (Eq.~\ref{eq:correlation1}) to highlight the growth of correlations over time.

Figures~\ref{f5_dex}~(a)-(b) show the growth of the second moment with time  defined in Eq.~\ref{eq:dex}. We observe different regimes of growth for $\Delta x(t)$ as well as for the return probability $\msr{R}(t)=\Pi(i=0,t)$~(see Figs.~\ref{f5_dex}(c)-(d)). The propagation of correlations measured via the saturation value of $\Delta x(t\rightarrow \infty)$ at very long time is increasing with $L$. This feature holds for both disorder strengths in the localized phase.
Moreover, $\Delta x(t)$ grows with a peculiar function, and thus not a clear algebraic either a logarithmic. This is possible due to the existence and competition of several time scales in the problem. For instance, in the non-interacting model, as mentioned before, typical non-interacting wavefunctions at $E \rightarrow 0$ decay as $\sim \exp(-\sqrt{|x|/\xi_0})$ rather than  exponentially. These states generate dynamical correlations that at low-energies give rise to extremely slow propagation~($t^{-1} \sim \exp(-\sqrt{|x|/\xi_0}) $) of correlations even in the non-interacting model~\cite{Krapivsky_2011, GDT_2016}. In the presence of interactions these states are mixing all over the spectrum but most likely are not completely destroyed by interactions; therefore, they give rise to some residual dynamics. Another possible time scale is the interaction strength, which is responsible for dephasing. 

Similarly, this behavior is also reproduced in the return probability $\msr{R}(t) = \Pi(i=0,t)$. As seen in Figs.~\ref{f5_dex}(c)-(d), for small $t$ the return probability seems to saturate. However, at longer times it again seems to decay with increasing system sizes. 

Although these time scales are hard to identify in the absence of an analytical technique, we provide a heuristic argument of such propagation in the low-density limit via  perturbation theory, which are described in the next section.  

\section{Two-interacting particles with off-diagonal disorder}\label{sec:Two_par}
In this section we aim to shed light on the observed subdiffusive dynamics. We demonstrate that the same type of dynamics occurs also for the case of only two particles which is always localized for any amount of disorder ($W>0$)~\footnote{At zero particle density short range interactions are not able to destroy a localized phase}.

Thus, let us start by inspecting the dynamics of the Hamiltonian $\msr{H}$ in the case in which the system hosts only two particles. In particular, we study the center of motion $\sigma_{+}^2(t)$ and of the difference $\sigma_{-}^2(t)$ of mass for two particles initialized next to each other  $|\psi \ra = c_{0}^\dagger c_{1}^\dagger |0\ra$ in the middle of the chain,
\eq{ 
\label{eq:center_of_mass}
\sigma_{+}^2(t) =  \sum_{i,j} \frac{(i+j)^2}{4} \la n_i n_j \ra (t), 
}
\eq{ 
\label{eq:diff_of_mass}
\sigma_{-}^2(t) =  \sum_{i,j} (i-j)^2 \la n_i n_j \ra (t).
}

\begin{figure}[tb]
\centering
 \includegraphics[width=1\columnwidth,keepaspectratio=true]{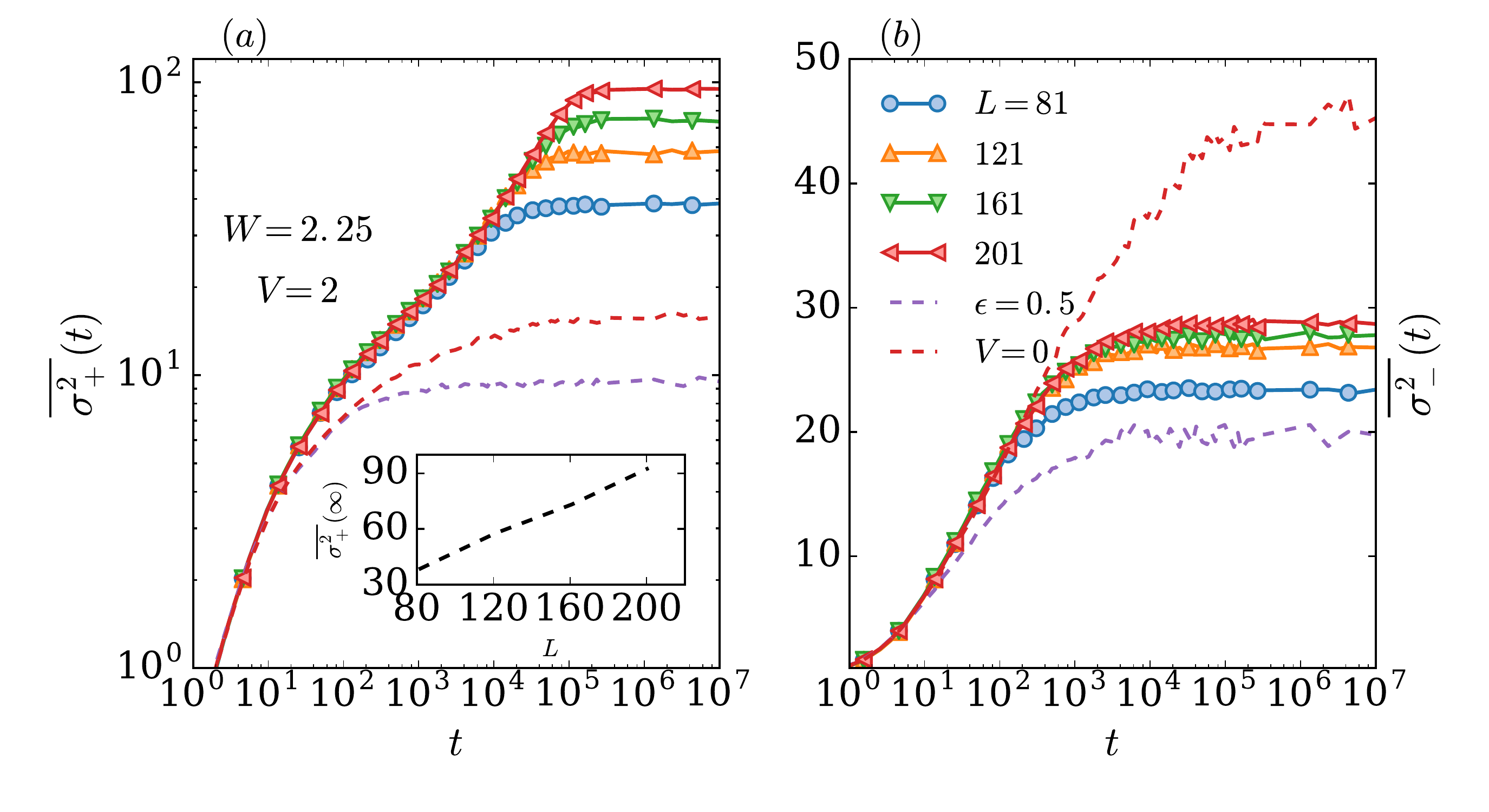}
\caption{(a): Algebraic growth of the center of mass $\overline{\sigma_{+}^2}(t)$ (Eq.~\ref{eq:center_of_mass}) of two interacting particles for $W=2.25$ and $V=2$. The inset shows the saturation value $\overline{\sigma_+^2}(\infty)$ for asymptomatic large times as a function of $L$, $\overline{\sigma_+^2}(\infty)\sim L$. (b): 
Difference of mass $\overline{\sigma_{-}^2}(t)$ (Eq.~\ref{eq:diff_of_mass}) of two interacting particles. In both panels the blue dashed lines $(L=201)$ depict the case in which a symmetry-breaking term has been added: $\msr{H}_\epsilon = \epsilon \sum_i \mu_i n_i$ with $\epsilon =0.5$ and $\{\mu_i\}$ random fields in $[-1,1]$. In both panels is also shown the non-interacting case $V=0$ for $L=201$. }
\label{f6_two}
\end{figure}

Figures~\ref{f6_two}~(a)-(b) show the dynamics of $\sigma_{+}^2(t)$ and $\sigma_{-}^2(t)$, respectively. The center of mass motion $\sigma_{+}^2(t)$ after the ballistic propagation at short times ($\sigma^2_+(t)\sim t^2$) shows a sub-diffusive growth. Importantly, at asymptotically large times $\sigma^2_+(t)$ saturates to a value, which scales linearly with $L$ ($\overline{\sigma_{+}^2}(\infty)\sim L$), as shown in the inset of Fig.~\ref{f6_two}~(a).
As a consequence, we expect that the propagation will be unbounded as $L\rightarrow \infty$. 
As we will show later, this propagation is a direct consequence of the PH symmetry of the non-interacting Hamiltonian. Indeed, breaking the symmetry by adding a small random potential $\msr{H}_{\epsilon} = \epsilon \sum_i \mu_i n_i$ with  $\{\mu_i\}$ random fields between $[-1,1]$, $\sigma^2_+(t)$ saturates quickly to an $L$-independent value, implying that the system is conventionally localized.

The spread of the difference of mass $\sigma_{-}^2(t)$ is fundamentally different. The two particles tend to remain close to each other, thus propagating coherently with a sub-diffusive relaxation of the center of mass. Moreover, Figs.~\ref{f6_two}~(a)-(b) show both the $\sigma_{+}^2(t)$ and the $\sigma_{-}^2(t)$ for the non-interacting case (dashed-lines, $V=0$). Comparing the interacting case with the non-interacting one, it is possible to notice that the effect of the interaction is to enhance only the propagation of the center of mass. At the same time the interaction bounds the motion of the two particles letting them propagate close to each other~\cite{Imry_1995, Shepe_94}. We emphasize that their bound state propagation is what we believe is the reason for having charge propagation in the MBL phase at finite particle density.

\begin{figure*}[tb]
\centering
 \includegraphics[width=1\textwidth,keepaspectratio=true]{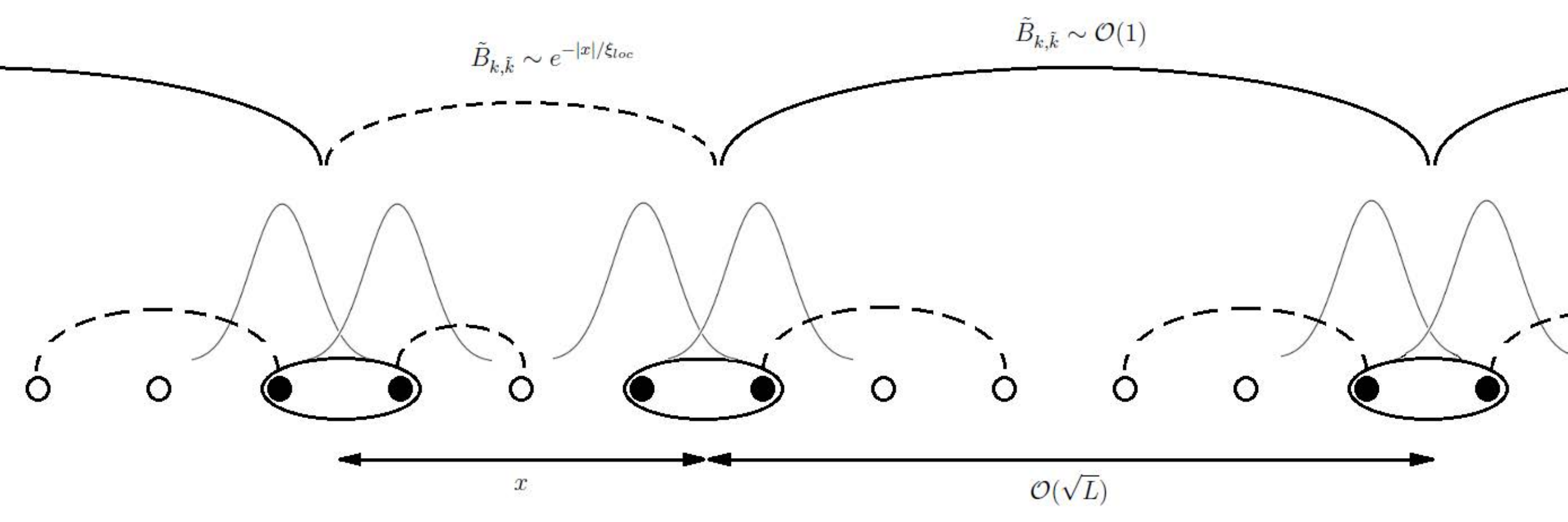}
\caption{Pictorial depiction of the coherent propagation two-particle states. Solid-lines represent the long-range hopping which link two-particle bound states at large distance. The dashed-lines are the exponentially suppressed hopping processes $\tilde{B}_{k, \tilde{k}}\sim e^{-|x_c(k) - x_c(\tilde{k})|/\xi_{\text{loc}}}$.}
\label{Fig_p}
\end{figure*}

\begin{figure}[tb]
\centering
 \includegraphics[width=1\columnwidth,keepaspectratio=true]{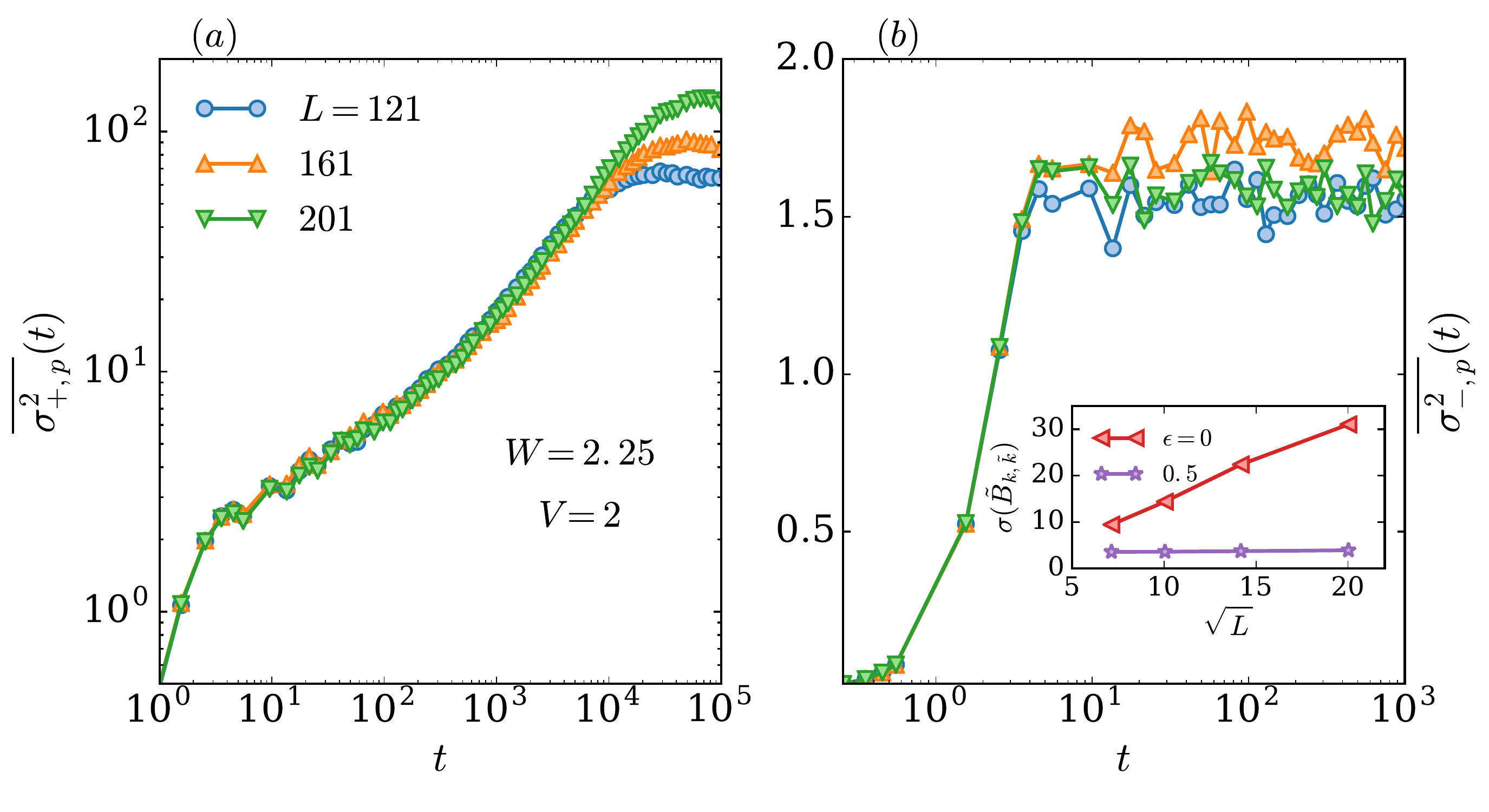}
\caption{(a): Center of mass $\overline{\sigma_{+,p}^2}(t)$. 
(b): Difference of mass $\overline{\sigma_{+,p}^2}(t)$.  The inset shows the typical hopping distance $\sigma (\tilde{B}_{k,\tilde{k}})$ (Eq.~\ref{eq:sigma_B}) given by the coupling $\tilde{B}_{k,\tilde{k}}$ in the effective Hamiltonian in Eq.~\ref{eq:H_eff_charge}. In presence of PH symmetry ($\epsilon =0$) 
$\sigma (\tilde{B}_{k,\tilde{k}})\sim \sqrt{L}$. Instead, by adding the symmetry-breaking term $\mathcal{H}_\epsilon$ with $\epsilon =0.5$, $\sigma (\tilde{B}_{k,\tilde{k}})\sim \mathcal{O}(L^0)$.
Both panels have been obtained restricting the dynamics onto the non-interacting eigenstates sub-space with total energy equal zero ($\eta^{\dagger}_k \eta^{\dagger}_{-k} |0\ra$). }
\label{f7_two}
\end{figure}

We now present an analytical argument supporting the sub-diffusive dynamics in the system, even though the system is localized. 
Initially, it is important to recall some basic properties of the non-interacting model $\msr{H}(V=0)$ and its single-particle eigenenergies $\{\epsilon_k\}$ and eigenfunctions $\{\psi_k(x)\}$. Due to the PH symmetry the single-particle eigenenergies come always in pairs $\epsilon_k$ and $\epsilon_{-k} = -\epsilon_{k}$. Furthermore, their respective eigenfunctions are related by $\psi_{-k}(x) = (-1)^x \psi_k(x)$. As a consequence $\psi_k(x)$ and $\psi_{-k}(x)$ are localized on the same center $x_c$ ($|\psi_k(x)|^2 = |\psi_{-k}(x)|^2 \sim e^{-2|x-x_c|/\xi_{\text{loc}}}$).
As we already mentioned the single-particle density of states is singular around zero and the eigenstates around the divergence are quasi-localized. For quasi-localized we mean that the wave-function is typically well localized at some center on the chain, nevertheless it has large fluctuations ($\sim \mathcal{O}(1)$) at a typical distance $\sim \sqrt{L}$ from the localization center~\cite{Ziman82, GDT_2016}. 

Moreover, we will make a small modification on the interaction term in $\msr{H}$ in Eq.~\ref{eq:ham} considering the two-body interactions $V\sum_i n_i n_{i+1}$ to better explain the mechanism of the dynamics.

Due to PH symmetry the non-interacting model has a degenerate eigen-spectrum. At half-filling this sector is composed by $\binom{L/2}{N/2}$ states and it is exponentially large in $L$. 
In particular, 
in the case of two particles, we will have $L/2$ states with total energy equal to zero ($\epsilon_k + \epsilon_{-k} =0$) of the form $\eta_{k}^\dagger \eta_{-k}^\dagger |0\ra$.  The states $\eta_{k}^\dagger \eta_{-k}^\dagger |0\ra$ are composed of two particles that are localized on the same center ($|\psi_k(x)| = |\psi_{-k}(x)|$). Thus, at strong disorder $W$ the initial state $|\psi \ra = c_{0}^\dagger c_{1}^\dagger |0\ra$ will have a large overlap with some states, which belong to the zero-energy sub-space. Moreover, at strong disorder and at finite interaction strength $V$, the states $\eta_{k}^\dagger \eta_{-k}^\dagger |0\ra$ are also close to the energy of the initial state $\la \psi |\msr{H} |\psi \ra \sim V$.

As first approximation we can restrict the dynamics to the two-particles sector composed by the non-interacting eigenstates $\eta_{k}^\dagger \eta_{-k}^\dagger |0\ra$ with total energy $E=0$, obtaining the following effective Hamiltonian
\eq{ \label{eq:H_eff_charge}
\msr{H}^{\text{eff}} = 2V \sum_{k, \tilde{k}} \tilde{B}_{k, \tilde{k}} \eta_k^\dagger \eta_{\tilde{k}} \eta_{-k}^\dagger \eta_{ -\tilde{k}},
}
where $\tilde{B}_{k,\tilde{k}} = \sum_{i} \psi_k(i) \psi_{\tilde{k}}(i)\psi_k(i+1)\psi_{\tilde{k}}(i+1)$. $\msr{H}^{\text{eff}}$ describes a hopping problem between the localized states $\eta_{k}^\dagger \eta_{-k}^\dagger |0\ra$'s. Most of the coupling terms $\tilde{B}_{k, \tilde{k}}$'s will couple the modes $k$ and $\tilde{k}$ only weakly, since $\tilde{B}_{k, \tilde{k}}$ involves the overlap of exponentially localized orbitals $\tilde{B}_{k, \tilde{k}}\sim e^{-|x_c(k) - x_c(\tilde{k})|/\xi_{\text{loc}}}$, they will not produce any substantial dynamics.  Nevertheless, the quasi-localized states close to the singularity of the single-particle density of state will allow long-range hopping of order $\sim \xi_{\text{loc}}(\epsilon_k \approx 0 )\sim \sqrt{L}$, producing thus a slow dynamics of the bound particles.

Figure~\ref{Fig_p} shows a  representation of the two particles dynamics trough the system. The solid-lines depict the long-range hoppings to a typical distance $\sim \mathcal{O}(\sqrt{L})$, while the dashed-lines are the exponentially suppressed hoppings $\tilde{B}_{k, \tilde{k}}\sim e^{-|x_c(k) - x_c(\tilde{k})|/\xi_{\text{loc}}}$.

To understand better the long-range hopping due to the existence of the quasi-localized modes, we analyze the coupling elements $\tilde{B}_{k,\tilde{k}}$ for $\epsilon_k\approx 0$. 
By detecting the center of localization $x_c(\tilde{k})$ for each single-particle eigenstates, we define 
\begin{equation}
\label{eq:sigma_B}
\sigma (\tilde{B}_{k,\tilde{k}}) = \overline{ \left (\frac{\sum_{\tilde{k}} |x_c(k)-x_c(\tilde{k})|^2 |\tilde{B}_{k,\tilde{k}}|}{\sum_{\tilde{k}}|\tilde{B}_{k,\tilde{k}}|} \right )^{1/2}},
\end{equation}
which quantifies the distance of the hopping between the two-particle states $(k,-k) \leftrightarrow (\tilde{k}, -\tilde{k})$. The inset of Fig.~\ref{f7_two}~(b) shows  $\sigma (\tilde{B}_{k,\tilde{k}})$ as a function of $L$, supporting the scaling $\sigma (\tilde{B}_{k,\tilde{k}})\sim \xi_{\text{loc}}(\epsilon_k \approx 0) \sim \sqrt{L}$, which also gives the right upper-bound for the scaling of $\sigma_+(\infty) \sim \sqrt{L}$. Moreover, as expected,  once the symmetry-breaking term $\mathcal{H}_\epsilon$ is added the effective Hamiltonian becomes short-range and $\sigma (\tilde{B}_{k,\tilde{k}}) \sim \mathcal{O}(L^0)$, also shown in the inset of Fig.~\ref{f7_two}~(b).

In order to support our assumption we rely on exact numerics, restricting the dynamics of the two particles to the non-interacting eigenstates with total energy equal to zero.  Figure~\ref{f7_two} shows the center $\sigma_{+,p}^2(t)$ and the difference $\sigma_{-,p}^2(t)$ of mass after projecting to the non-interacting sub-space with $E=0$. $\sigma_{+,p}^2(t)$ shows the same sub-diffusive dynamics as in the case without the projector, giving thus evidence that this propagation is due to the interplay between interactions and the non-interacting sub-space with $E=0$. 

Moreover, $\sigma_{-,p}^2(t)$ saturates to an $L$-independent value as a consequence of projecting on the non-interacting eigenstates in which the two-particles are localized on the same center (see also Fig.~\ref{f6_two}~(b)).

\section{Conclusion}
In summary, we investigated the dynamics, i.e., entanglement and charge propagation, in a particle-hole symmetric MBL phase. In particular, we studied the $t{-}V$ model with random bonds (off-diagonal disorder). The non-interacting limit of this model is known to have a divergence in the single-particle density of states at zero energy. As a consequence, the single-particle modes are quasi-localized at energies close to zero, while at other energies states are exponentially localized.

We provide numerical evidence of the stability of the localized phase at strong disorder once interactions are switched on, by studying several observables that establish the MBL phase. For instance, at sufficiently strong disorder, the level-statistics of energies is Poissonian, the probability distribution of the bipartite entanglement entropy is highly non-thermal and presents the typical exponentially decaying tails of a localized phase. Nevertheless, it is important to point out that for this particular model we found strong finite-size effects in these observables and in particular, we cannot rule out the possibility of logarithmic corrections in system size for the scaling of the entanglement entropy as found in SU(2) symmetric models.


Next, we characterized the information propagation through the system by studying the time evolution of the entanglement entropy after a quantum quench. At weak interactions, we employed a recently proposed method to study information propagation in the MBL phase for large time scales. We found that the entanglement entropy grows logarithmically in time, as suggested from a strong disorder renormalization group calculation.
Nevertheless, at stronger interactions but still in the MBL phase, the growth is consistent with an algebraic propagation of information with a non-thermal saturation for asymptotically long times.

Finally, we studied the charge propagation through the system by employing the density-density correlator at infinite temperature. We detect a slow propagation of charge for the considered time scales, even though the system can be thought as `conventionally' localized. We presented an analytical argument corroborated with exact numerics, based on two-particles propagation, explaining the main phenomenology of this anomalous transport.

In conclusion,  our results suggest that entanglement and charge propagation is different in PH symmetric system from a conventional MBL phase (without any symmetry). For the later case no charge propagation is expected and the entanglement grows only logarithmically in time. 

\section{ACKNOWLEDGMENTS}
We would like to thank R. Bhatt, W. De Roeck, F. Evers, K. H\'em\'ery , A. D. Mirlin and P. Sala for several discussions. 
We  also  express  our  gratitude  to P. Sala for a critical reading of the manuscript.
SB acknowledges support from DST, India, through Ramanujan Fellowship Grant No. SB/S2/RJN-128/2016,  through early career award ECR/2018/000876, and MPG for funding through the Max Planck Partner Group at IITB. MH acknowledges support by the Deutsche Forschungsgemeinschaft via the Gottfried Wilhelm Leibniz Prize program. GDT and SB acknowledge the hospitality of MPIPKS Dresden where part of the work was performed.

\bibliography{OFF}

\end{document}